\documentstyle[a4]{article}

\begin{document}

\noindent {\bf Comment on "Vortex Liquid Crystal in Anisotropic Type II Superconductors"}  

\vskip5mm

Recently Carlson et al.  \cite{Carlson} tried to compose the \(H-T\) phase diagram of
anisotropic superconductors based on the Lindemann criterion for melting of 
flux-line lattice. It is worthy to notice that the issue itself is interesting.  However,
their conclusion on the existence of a smectic vortex phase is unphysical for the following
reasons: There is no appropriate estimate on thermal fluctuations above the 
"lower melting temperature", since the basis of the elastic theory 
used by the authors, namely the lattice structure, disappears already; 
One therefore cannot argue even the existence of the intermediate phase, such as smectic
in Fig. 1 of Ref.\cite{Carlson},
above the "lower melting temperature", not only its melting as noticed by the authors. 

It is interesting to ask if one can draw useful information from a Lindemann-type
argument for melting phenomena in uniaxially anisotropic superconductors when
the magnetic field is applied perpendicularly to the anisotropic axis.  
An idea is to suppose that one can have {\it two Lindemann numbers} in the two different directions,
instead of two meltings.  From the elastic theory (see references in Ref.\cite{Carlson}), 
we know that the Lindemann melting line is given by the following equation:
\(t/\sqrt{1-t}=c^2\times f(b,\kappa,\gamma,Gi)\).  This property permits 
us to shift the "upper melting line" in Fig. 4 of Ref.\cite{Carlson} to lower temperature 
by reducing the Lindemann number without touching the details of \(f(b,\kappa,\gamma,Gi)\)
\cite{Note}. The numerical results thus obtained are shown in Fig. 1.  There
is a perfect collapse between the two curves when one takes \(c_{long}\simeq 0.17\) while
fixes \(c_{short}=0.2\) to the value chosen in Ref.\cite{Carlson}.  The collapse of the two curves
is expected for large \(\kappa\)'s, namely in extremely type II superconductors, and is achieved
whenever the two Lindemann numbers satisfy the fixed ratio which is revealed in our analysis.  

Should the "two melting lines" never collapse, there appears another possible melting scenario.
One could have an intersection between
these "two melting lines", besides the ones at \(T=T_c\), \(H=0\) and \(T=0\), \(H=H_{c2}\).
Then, the lower melting temperature should be adopted as the true one at a given
magnetic field to form a single melting line.  The "intersecting point" could be a triple 
point: the melting mechanism below and above the triple 
magnetic field \(H_t\) is different; there may be two liquids separated by another first-order 
line, which terminates at a critical point where the two liquids become identical.

There exists of course the possibility of an intermediate phase in the \(H-T\) phase
diagram (see for example \cite{Hunew}). Its identification is beyond the 
Lindemann-type phenomenological argument. 
Although terminologies of superconductivity have been used in the present discussions,
it is obvious that the results are applicable to a wide range of melting phenomena.

This study is partially supported by the Ministry of Education, Culture, Sports, Science and 
Technology, Japan, under the Priority Grant No. 14038240.

\vskip10mm

\noindent Xiao Hu and Qing-Hu Chen\\
\noindent Computational Materials Science Center \\ 
\noindent National Institute for Materials Science \\
\noindent Tsukuba 305-0047, Japan

\vskip5mm

\noindent Received March 5 2003

\vskip5mm

\noindent PACS numbers: 74.25.Op, 61.30.-v, 74.25.Qt


\vskip10mm

\noindent Figure caption:

\noindent Fig. 1: Collapse of the two curves in Fig. 4 of Ref.\cite{Carlson}
by tuning the Lindemann number in the "long" direction to 0.17 while keeping 
that in the "short" direction at 0.2.


\begin{thebibliography}{9}
\bibitem{Note} The authors of Ref.\cite{Carlson} did not provide
enough information, such as \(T_c\) or \(Gi\), for one to reproduce their Fig. 4.
\end{thebibliography}
\end{document}